**Improving Models for Student Retention and Graduation using Markov Chains**

**Authors:**

Mason N. Tedeschi[1], Tiana M. Hose[2], Emily K. Mehlman[3], Scott Franklin[4], Tony E. Wong[2]*

**Affiliations:**

[1] New College of Florida, Sarasota, FL 34243

[2] School of Mathematical Sciences, Rochester Institute of Technology, Rochester, NY 14623

[3] College of Science, Rochester Institute of Technology, Rochester, NY 14623

[4] School of Physics and Astronomy, Rochester Institute of Technology, Rochester, NY 14623

* Corresponding author: aewsma@rit.edu

**Abstract:**

Graduation rates are a key measure of the long-term efficacy of academic interventions. However, challenges to using traditional estimates of graduation rates for underrepresented students include inherently small sample sizes and high data requirements. Here, we show that a Markov model increases confidence and reduces biases in estimated graduation rates for underrepresented minority and first-generation students. We use a Learning Assistant program to demonstrate the Markov model's strength for assessing program efficacy. We find that Learning Assistants in gateway science courses are associated with a 9% increase in the six-year graduation rate. These gains are larger for underrepresented minority (21%) and first-generation students (18%). Our results indicate that Learning Assistants can improve overall graduation rates and address inequalities in graduation rates for underrepresented students.

# 1 Introduction

University decision-makers invest in educational programs to support the success of diverse students. Assessing programmatic efficacy, however, is made complicated by complexities in program implementation and measurement of program outcomes. Administrators often measure programmatic efficacy using a single metric of student success, such as course failure rate (commonly rates of D, F, or withdrawal; DFW), and make decisions about funding or program longevity using this information. While this strategy provides an important view of one aspect of program efficacy, the analysis is limited to a single point in a student's academic career. This limitation risks missing other important student success outcomes across longer timescales, including retention and graduation rates, that the educational program may impact. Thus, there is a need for statistical methods to incorporate measures of longer-term, or "downstream", student success into program evaluation metrics.

Additionally, if an educational program has a small scope (e.g., implementation in a single section of a course) or administrators are interested in the program's impacts on students from minoritized identities (e.g., women in science majors or underrepresented minority students), then program assessment is challenged by small sample sizes. Smaller sample sizes lead to lower confidence in measures of student success such as DFW or retention rates. This creates a need for methods to estimate more accurate statistics for program efficacy metrics with small sample sizes.

When program managers and administrators use graduation rates as a measure of program success, there is an inherent discrepancy between the students who are currently engaged in the particular program under study, and the students whose data are part of the graduation rate calculation. Specifically, very few of the students who are currently in an academic program are involved in a "current" estimate of the program's graduation rate. Most of these students have departed the program, with or without a degree (or other marker of successful completion of the program). Thus, there is a benefit to developing an "online" estimate of graduation rates - that is, an estimate of graduation rates that can be updated regularly [1] and includes data from all cohorts of students, including those who are still enrolled in the program. Such an online estimate of graduation rate is arguably more appropriate for program evaluation in that it makes use of the most up-to-date information on student retention and persistence. A traditional estimate of six-year graduation rate (SYGR), on the other hand, relies entirely on data that is at least six years old. If, five years ago, there was a sizable improvement in first-year retention in the academic program under study, then the six-year graduation rate as estimated in the following year will be higher as a result (all other things being equal). This highlights the strength of the online estimate, in that it can incorporate all available information on year-to-year transitions.

As Markov chains model transitions of a system between states, they provide a natural mathematical structure to represent the year-to-year transitions of students between academic states (e.g., freshman year, sophomore year). A (first-order) Markov model is one in which the probability of the system transitioning from one state to another is conditionally independent of all previous states of the system, given the current state. Previous work has leveraged Markov chains to construct estimates of college students' retention and persistence [1–3]. Boumi and Vela [1] further allowed for reflexive transitions from a state to itself; that is, students can remain in the same academic level at multiple points in time. This "multi-level absorbing Markov chain" allowed multiple visits to a state either in sequence (e.g., a student remaining in the sophomore state for more than one academic year) or non-contiguous (a student withdrawing for multiple, non-contiguous semesters then re-enrolling). Here, the term "absorbing" refers to the fact that once a student enters either the Graduation state or the Drop-Out state, they are assumed to remain in that state. Thus, in the parlance of Markov chains, these system states are said to be absorbing states. By contrast, each academic year is a transient state because students pass through these states en route to a permanent outcome (i.e., one of the two absorbing states). However, the introduction of reflexive transitions by Boumi and Vela [1] leads to an underestimation bias in Markov-based estimates of SYGR. Those authors develop a correction to account for this bias, but to avoid this issue, we focus on graduation within six calendar years of students' matriculation.

Here, we assess the ability of a Markov chain model to constrain estimates of SYGR by comparing its performance to that of a traditional graduation rate calculation. We use eight years of data from Rochester Institute of Technology (RIT), a large, private university in Rochester, New York, United States, to compute SYGR estimates and we characterize uncertainty by constructing bootstrap confidence intervals. Specifically, we hypothesize that the use of more data in the Markov estimates will lead to more confident estimates of SYGR than a traditional calculation. This ability to use more data makes an online estimate of graduation rates suitable for small groups of students. We demonstrate the power of the analysis by using it on a localized academic program, the *Learning Assistant* (LA) program, and smaller demographic sub-groups within the program. However, we stress that an online estimate of graduation rates has no drawback relative to a traditional calculation. The online estimate will make use of all of the same information as a traditional estimate, as well as the year-to-year transition information from not-yet-departed students. This makes the online estimate appropriate for other settings as well. We assess, through our real-world data from RIT, the benefits of using the Markov model to estimate SYGR for smaller, minoritized groups of students (underrepresented minorities and first-generation college students). Finally, using the LA program at RIT as a test case, we evaluate the impacts of students' exposure to LAs in terms of improvements to SYGR for students who have experienced LA support in a high-attrition science course.

## 2 Methods

### 2.1 Data and Context

We analyze the academic progress of students at RIT, the second largest producer of STEM graduates among all private universities in the United States. RIT is a primarily undergraduate institution with approximately 15,000 undergraduate and 3,000 graduate students. The student body is rich in diversity, with more than 2,500 international students and 2,900 African American, Latino American, and Native American (AALANA) students. RIT hosts the National Technical Institute for the Deaf, with over 1,500 deaf/hard-of-hearing students taking courses alongside their hearing peers. Data for this study includes *all* RIT students, and comes from the RIT Office of Institutional Data. The data include previous and subsequent academic records, program (major) at the time of each class, and, if the student graduates, the graduation year and degree obtained.

We demonstrate the power of Markov analysis by using it to evaluate the effectiveness of a localized program, RIT's *Learning Assistant* (LA) program. In this program, RIT hires undergraduate Learning Assistants (LAs) to support instructors using evidence-based, student-centered learning practices in their courses. LAs receive training in pedagogical knowledge and pedagogical content knowledge through planning sessions with faculty and a pedagogy course [4,5] to support them as effective near-peer educators in the classroom. STEM faculty use LAs to implement evidence-based teaching strategies known to increase learning [6,7] and lower DFW rates [7–10]. A scoping review [11] of 39 peer-reviewed studies found that LA-supported courses were associated with improvements in higher-order cognitive skills [12] and improved conceptual learning [6], a result that did not depend upon a single set of curricular materials (e.g., [13–15]). DFW rates in LA-supported courses are lower than in non-supported courses [7,9,10], with increased impacts on students from underrepresented demographic groups [8]. Importantly, previous work has not closely examined how LA support relates to longer-term metrics such as graduation and retention rates, nor has a direct comparison between concurrent supported and unsupported classes been possible. At RIT, the LA program is implemented in a subset of sections of particular courses, and Markov analysis enables a comparison between the performance of the LA-supported students and the "control" group of students in unsupported sections. In this way, RIT's LA program naturally gives rise to contexts in which improved program efficacy metrics for small data situations could provide decision-makers with more confident estimates of student success.

## 2.2 Traditional estimate of six-year graduation rate

Traditional estimates of graduation rates require first identifying a student cohort of interest. Suppose $N_{start}$ is the initial number of students who matriculate in that cohort. Among those students, let $N_{deg}$ represent the number of students who obtain a degree within six years. Then $GR_{tr} = N_{deg}/N_{start}$ is the estimate of the SYGR for this cohort of interest. A key limitation of using this traditional approach is that $N_{start}$ is frequently quite small when the cohort of interest is an underrepresented group of students, or a relatively small subgroup of students who experience an academic intervention.

## 2.3 Markov chain-based estimate of six-year graduation rate

### 2.3.1 Model

In contrast to the traditional SYGR calculation, we employ a Markov model that estimates the SYGR by leveraging the probabilities of students persisting from year-to-year, conditioned on their initial class status (e.g., freshman, sophomore).

Define S = {1, 2, 3, 4, 5, 6, D, G} as the state space for our Markov chain model, where the integers 1, 2, …, 6 represent the numbers of years that a student has been enrolled at RIT (e.g., students in state 2 are generally sophomores), D represents students who have dropped out and no longer are pursuing a degree, and G represents students who have earned a degree. States D and G are both absorbing states, while the others are transient.

Suppose that $N_i$ represents the number of students in state i at the beginning of a year, and let $N_j$ represent the number of those students who are in state j by the end of the year. Then the transition probability from state i to state j is

$$P_{ij} = P(X_{t+1} = j \mid X_t = i) = N_j/N_i. \qquad (Eq.\ 1)$$

This probability is by definition 0 if j is not in {i+1, D, G}. Additionally, by definition, $P(X_t+1 = j \mid X_t = D) = 0$ if j not equal to D, and similarly $P(X_{t+1} = j \mid X_t = G) = 0$ if j not equal to G. Thus, the overall structure of the transition probability matrix is sparse (Table 1):

*Table 1. General structure of the transition probability matrix for state space {1, 2, 3, 4, 5, 6 (year at university), D (Drop out), G (Graduate)} to represent the conditional probability of the system transitioning to a final state (column), given that the system starts in a particular initial state (row).*

|   | 1 | 2 | 3 | 4 | 5 | 6 | D | G |
|---|---|---|---|---|---|---|---|---|
| 1 | 0 | $P_{12}$ | 0 | 0 | 0 | 0 | $P_{1D}$ | $P_{1G}$ |
| 2 | 0 | 0 | $P_{23}$ | 0 | 0 | 0 | $P_{2D}$ | $P_{2G}$ |
| 3 | 0 | 0 | 0 | $P_{34}$ | 0 | 0 | $P_{3D}$ | $P_{3G}$ |
| 4 | 0 | 0 | 0 | 0 | $P_{45}$ | 0 | $P_{4D}$ | $P_{4G}$ |
| 5 | 0 | 0 | 0 | 0 | 0 | $P_{56}$ | $P_{5D}$ | $P_{5G}$ |
| 6 | 0 | 0 | 0 | 0 | 0 | 0 | $P_{6D}$ | $P_{6G}$ |
| D | 0 | 0 | 0 | 0 | 0 | 0 | 1 | 0 |
| G | 0 | 0 | 0 | 0 | 0 | 0 | 0 | 1 |

The six-step transition probability of a student arriving in the graduation state (G) within six years of matriculation (originating in state 1) are $P^6_{1G} = P(X_{t+6} = G \mid X_t = 1)$. These provide an estimate of SYGR: $GR_{ma} = P^6_{1G}$, the row 1, column G (eighth column) of the sixth power of the transition probability matrix P.

A key strength of the Markov modeling approach is the ability to combine information from multiple cohorts of students. For example, let A and B represent two cohorts of interest. Let $N_{A,i}$ and $N_{B,i}$ represent the numbers of students from cohorts A and B who are in state i at the beginning of a year. Similarly, let $N_{A,j}$ and $N_{B,j}$ represent the numbers of students who are in state j by the end of that year. Then the combined estimate of the transition probability from state i to state j is

$P_{AB,ij} = (N_{A,j} + N_{B,j})/(N_{A,i} + N_{B,i})$. (Eq. 2)

Using Equation (2) enables the Markov model to include the most up-to-date data . For example, as of this writing, students who matriculated in Fall 2021 have only been at university for one full year. Using traditional methods, a SYGR estimate that includes these students will not be available for another five years. Their data, however, can be incorporated into the Markov model for SYGR by influencing the estimates of $P_{12}$, $P_{1D}$, and $P_{1G}$. For such a "partial data" cohort, they do not influence any of the subsequent transition probability estimates.

### 2.3.2 Validation

We first perform a "positive control" experiment to demonstrate that a reduced form of the Markov model produces results that match the traditional SYGR calculation. For each of the cohorts matriculating in Fall 2013, Fall 2014, and Fall 2015, we construct a reduced form of the Markov model by only using the Fall 2013, 2014, or 2015 cohorts' data (respectively) to construct each Markov transition probability matrix (Eq. 1). In this way, these models employ the same information that traditional estimates of SYGR use for those same cohorts. We compare the graduation rate estimates from these reduced form Markov models to traditional graduation rate estimates as a way to validate that the Markov model faithfully reproduces the traditional baseline.

In contrast to this positive control, the full form of the Markov model involves a single full data cohort, and five partial data cohorts. For example, a program manager in Fall 2021 could use the full six years of data for the Fall 2015 cohort in order to estimate the SYGR using the traditional approach. They could also employ the Fall 2016, 2017, 2018, 2019, and 2020 cohorts' information in an estimate of SYGR using the Markov chain approach. Further, the Markov chain model provides a natural structure (Eq. 2) to combine all of the available information into a single estimate of SYGR by also including the Fall 2013 and 2014 cohorts' data. This constitutes the "full" Markov model for SYGR.

### 2.4 Bootstrapping uncertainty characterization

To test our hypothesis that a Markov model will lead to more confident estimates of graduation rates than a traditional model, we quantify uncertainty in our estimates of SYGR by constructing confidence intervals using bootstrap resampling. We do this for both the traditional and the Markov chain models. We resample with replacement rows from the original overall student data set, resampling a number of rows equal to the original size of the data set. For each resample, we compute the SYGR using the traditional (or Markov) approach. We use ensembles of 1,000 resamples and estimate a 95% confidence interval as the 2.5-97.5% range in the ensemble of computed SYGR. Experiments using different sizes of resample ensembles indicate this ensemble size is sufficiently large that our results are not sensitive to this choice.

## 2.5 Subsets of students

We demonstrate the strength of the Markov model for SYGR for studying relatively small cohorts of students by examining how SYGR differs for students in courses supported by undergraduate LAs and those without such LA support. Further, we examine how this impact of LA support may have differential effects for underrepresented minority students and first generation college students. All of these groups (LA-supported, underrepresented minority, and first generation) are relatively small on their own, so their intersections provide an ample proving ground for the benefits of the Markov model for SYGR (e.g., LA-supported first generation college students). For these experiments using smaller subgroups of students, we do not present any results using the traditional approach to compute SYGR because the benchmarking experiment (Sec. 2.3.2) will demonstrate the fidelity with which the Markov model estimates SYGR.

To control for the variety of students' major disciplines, and the variety of courses that LA support is incorporated into, we include the LA-supported group only students who encounter LA support in high-attrition courses within the College of Science at RIT. These include introductory physics and mathematics courses, such as precalculus and calculus (see Supplementary Material). We note that other high-attrition courses are present, but few or no sections that incorporate LAs. For these experiments, we restrict our data set to only students who take at least one of these courses between Fall 2013 and Summer 2021.

Students who encounter LA support in at least one of the aforementioned courses contribute to the Markov-based calculation starting in the first year in which they have LA support. In this way, the transition probabilities in the Markov transition matrix for the LA-supported students should only represent transitions which would have been affected by the LA support "treatment". By contrast, the "control" group of students includes all students who have no exposure to LA support in the high-attrition science courses (see above). The numbers of students and class sections each semester, and the breakdown between LA-supported and non-LA-supported, is provided in Supplementary Material.

We note that separating students based on LA status (LA-supported or not) and examining the groups' differences in graduation rates establishes a correlative or associative relationship, but cannot prove a causal one. This is, of course, the nature of all such treatment-versus-control type experiments (e.g., vaccine trials). The deep existing literature documenting the benefits of LA support for students across different institutional profiles and disciplines – including transferable and durable gains in higher-order cognitive skills, conceptual learning, and problem-solving skills (see Sec. 1) – provides strong reason to hypothesize that students' exposure to LA support should be related causally to their likelihood of graduation within six years.

## 3 Results

### 3.1 Benchmarking against traditional graduation rate calculation

When only cohorts with a full six years of data are used to compute SYGR using the Markov method, estimates of $GR_{ma}$ from the reduced form Markov model faithfully match the traditional graduation rate estimates, $GR_{tr}$ (Table 2). To within a percentage point, the medians and 2.5-97.5% bootstrap confidence interval bounds (and widths) are all consistent for the Fall 2013,

Fall 2014, and Fall 2015 cohorts, as well as when all three cohorts' data are combined (using Eq. 2). These results provide strong evidence that the Markov model is a valid approach to estimate SYGR. For the remainder of this work, the Markov model uses all student data that would be hypothetically available at the time when a traditional SYGR estimate could be computed. For example, the Markov-based SYGR for the Fall 2013 cohort will include five years of data for the Fall 2014 cohort, four years of data for the Fall 2015 cohort, and so forth. The "all combined" model includes all available student data in our data set from Fall 2013 to Summer 2021 (the last academic term in our data set). We use "percentage point" to refer to ranges and absolute differences in the SYGR to avoid confusion with relative changes or relative differences in SYGR.

*Table 2.* Estimates of six-year graduation rate (%) from the traditional calculation and the reduced form (full-data cohorts only) Markov model. Numbers are rounded to the nearest percentage point, so the 95% CI width reported may not appear to equal the difference between the 97.5th and 2.5th percentiles.

| Cohort | Method | 2.5th percentile | Median | 97.5th percentile | 95% CI width |
|---|---|---|---|---|---|
| Fall 2013 | Traditional | 66 | 71 | 75 | 8 |
| Fall 2013 | Markov | 66 | 71 | 75 | 9 |
| Fall 2014 | Traditional | 66 | 71 | 75 | 9 |
| Fall 2014 | Markov | 67 | 71 | 75 | 9 |
| Fall 2015 | Traditional | 69 | 73 | 77 | 9 |
| Fall 2015 | Markov | 69 | 73 | 77 | 8 |
| All combined | Traditional | 69 | 72 | 74 | 5 |
| All combined | Markov | 69 | 72 | 74 | 5 |

## 3.2 Quantification of uncertainty

The Markov model produces estimates of SYGR that are substantially more confident than the estimates of SYGR using the traditional approach (Fig. 1). For the Fall 2013-2015 cohorts, using the Markov method results in 95% CI widths that are tightened by between 33% and 44% relative to the CI widths following the traditional calculation. Importantly, all of the median SYGR estimates are consistent across the two methods to within 1 percentage point: 71% for Fall 2013 for both methods; 71% for Fall 2014 for both methods; and 73% for Fall 2015 using the traditional approach, as compared to 72% when using the Markov model (Table 3). When using all available data, the reduction in 95% CI width is somewhat lower (34% relative to the traditional calculation) but still appreciable. This is attributable to the fact that using all available data means that the amount of information from partial-data cohorts (Fall 2016 to present) is relatively smaller compared to the full-cohort information (Fall 2013-2015).

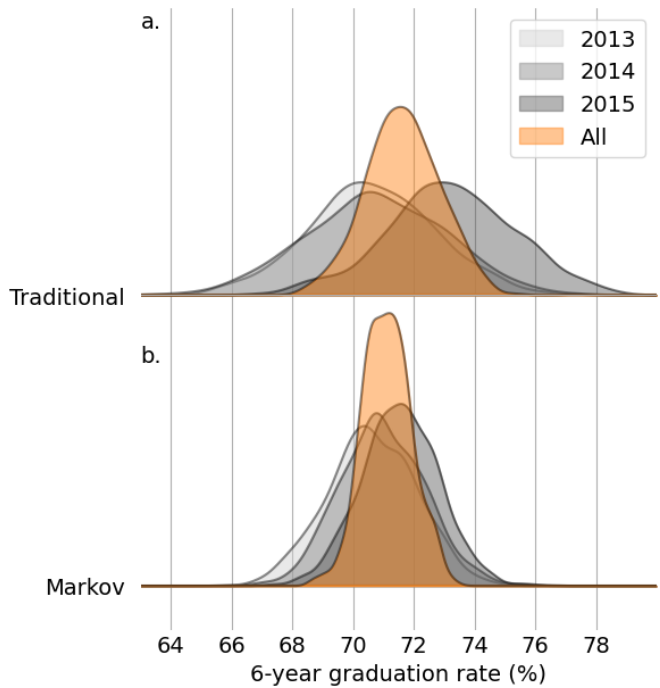

*Figure 1.* Using a Markov model to estimate SYGR (bottom panel) leads to tighter 95% confidence intervals than when a traditional approach is used (top panel). Shown are kernel

*density estimates for six-year graduation rate for the Fall 2013 (light gray), Fall 2014 (gray), and Fall 2015 (dark gray) cohorts and all three of these cohorts combined (orange).*

*Table 3. Estimates of six-year graduation rate (%) from the traditional calculation and the full-form Markov model. Numbers are rounded to the nearest percentage point, so the 95% CI width reported may not appear to equal the difference between the 97.5th and 2.5th percentiles.*

| Cohort | Method | 2.5th percentile | Median | 97.5th percentile | 95% CI width |
|---|---|---|---|---|---|
| Fall 2013 | Traditional | 66 | 71 | 75 | 8 |
| Fall 2013 | Markov | 68 | 71 | 73 | 6 |
| Fall 2014 | Traditional | 66 | 71 | 75 | 9 |
| Fall 2014 | Markov | 68 | 71 | 74 | 6 |
| Fall 2015 | Traditional | 69 | 73 | 77 | 9 |
| Fall 2015 | Markov | 69 | 72 | 74 | 5 |
| All combined | Traditional | 69 | 72 | 74 | 5 |
| All combined | Markov | 70 | 71 | 73 | 3 |

### 3.3 Small subsets of students

As expected, the CI widths for smaller subsets of students (AALANA students and first-generation college students) are notably wider than for the general student population (Table 4). This is due to the fact that the sample sizes for these underrepresented groups are smaller, by their nature as underrepresented groups. Importantly, similar benefits in terms of reduced CI widths are seen in the Markov-based estimates of SYGR for AALANA students and first-generation students in College of Science majors (Table 4).

*Table 4.* *Estimates of six-year graduation rate (%) for AALANA (top set) and first-generation (bottom set) science majors, using both the traditional calculation and the full-form Markov model. Numbers are rounded to the nearest percentage point, so the 95% CI width reported may not appear to equal the difference between the 97.5th and 2.5th percentiles.*

|  | Cohort | Method | 2.5th percentile | Median | 97.5th percentile | 95% CI width |
|---|---|---|---|---|---|---|
| **AALANA** | Fall 2013 | Traditional | 40 | 55 | 68 | 28 |
|  |  | Markov | 54 | 62 | 71 | 17 |
|  | Fall 2014 | Traditional | 48 | 59 | 70 | 22 |
|  |  | Markov | 57 | 64 | 73 | 17 |
|  | Fall 2015 | Traditional | 59 | 70 | 83 | 24 |
|  |  | Markov | 62 | 69 | 77 | 15 |
|  | All combined | Traditional | 55 | 62 | 69 | 14 |
|  |  | Markov | 61 | 65 | 70 | 9 |
| **First-generation** | Fall 2013 | Traditional | 61 | 70 | 81 | 19 |
|  |  | Markov | 64 | 70 | 76 | 11 |
|  | Fall 2014 | Traditional | 57 | 67 | 77 | 20 |
|  |  | Markov | 61 | 68 | 76 | 15 |
|  | Fall 2015 | Traditional | 63 | 72 | 81 | 17 |
|  |  | Markov | 65 | 70 | 75 | 10 |
|  | All combined | Traditional | 64 | 70 | 75 | 11 |
|  |  | Markov | 67 | 70 | 73 | 7 |

The Markov approach improves the confidence interval widths by between 25 and 42% (tightening) relative to the traditional model (Figure 2a and c, and Table 4). It is notable that the CIs for AALANA science majors using the traditional approach for computing SYGR are consistently more than 20 percentage points wide, with sizable variation from year-to-year in the median estimate (55% in Fall 2013, 59% in Fall 2014, and 70% in Fall 2015). This interannual variability when the traditional approach is used is problematic, as students and academic program managers may make decisions based on SYGR information that varies widely from year to year. For example, when using the Fall 2013 cohort, the 95% CI for AALANA science major SYGR spans from 40 to 68%, which does not even include the median SYGR when using the Fall 2015 cohort (70%). While this does not necessarily show statistically significant differences from year to year, these results highlight that for small subsets of students, there are substantial variations in best estimates of SYGR when using the traditional approach.

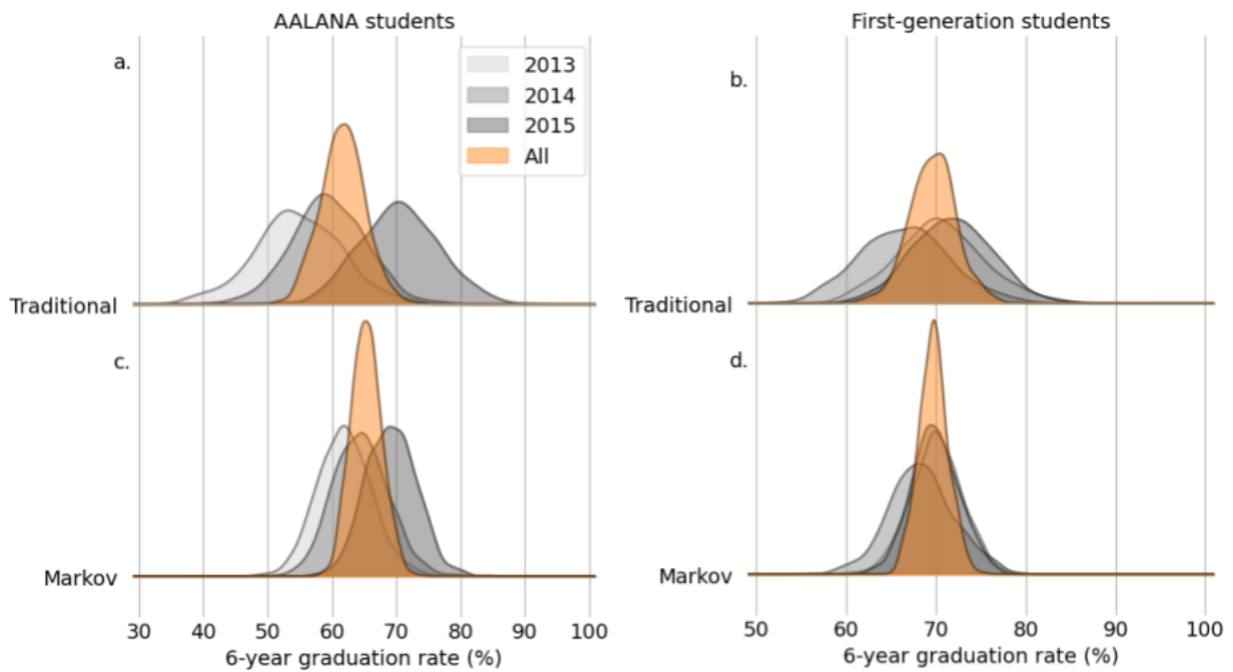

*Figure 2.* Using a Markov model (bottom panels) to estimate SYGR for AALANA (left panels) and first-generation (right panels) science majors leads to tighter 95% confidence intervals and decreases year-to-year variability as compared to a traditional approach (top panels). Shown

*are kernel density estimates for six-year graduation rate for the Fall 2013 (light gray), Fall 2014 (gray), and Fall 2015 (dark gray) cohorts and all three of these cohorts combined (orange).*

By contrast, using the Markov model incorporates more information as it becomes available. Thus, the estimates of SYGR are better centered on the long-term median: 62% in Fall 2013, 64% in Fall 2014, 69% in Fall 2015, and 65% overall, for AALANA students. Further, the CI widths are consistently below 20 percentage points. Similar gains in terms of tighter CIs and lower interannual variability in estimated SYGR are seen when the Markov model is applied to the subset of first-generation science majors (Figure 2b and d, and Table 4).

### 3.4 Proof of concept: Evaluating the impacts of LA support

We focus now on students who, at some point during their time at RIT, were enrolled in at least one of the high-attrition science courses and had declared a science major (see Sec. 2.5). We separate this group of students into two groups: the "LA group" of students had an undergraduate Learning Assistant support at least one of their high-attrition science courses; the "no-LA group" were enrolled in sections of these high-attrition courses that were not supported by an LA. All results presented in this section include all available student data from Fall 2013 to Summer 2021 (the "all combined" case from the previous Results sections), and employ the Markov model.

We find that exposure to LAs in high-attrition science courses is associated with a 9 percentage point improvement in SYGR, relative to students who were enrolled in sections of these courses that did not have LA support (Figure 3a). LA-supported students have a median SYGR of 77% (95% CI of 73-81%). This is statistically significantly higher than the non-LA group, which has a median SYGR of 68% (95% CI of 66-70%).

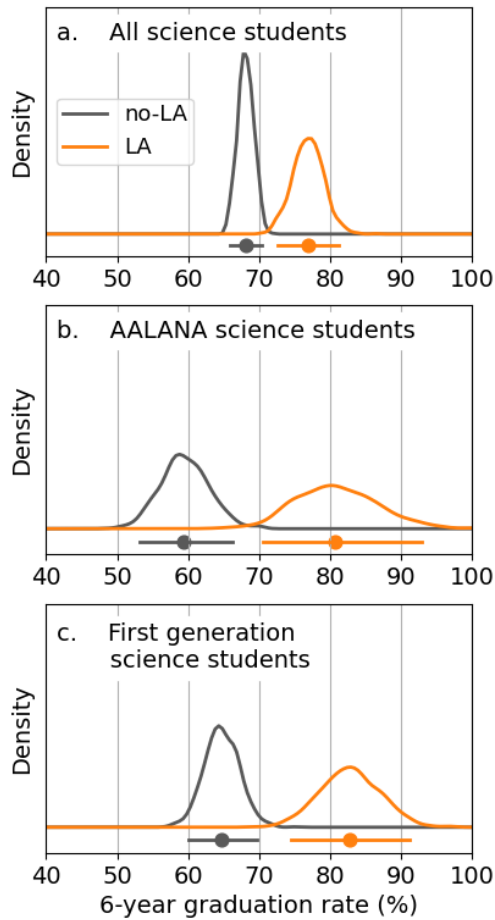

*Figure 3.* *Learning Assistant support is associated with a 9 percentage point improvement in six-year graduation rates for science majors in general (a). These gains are even larger for AALANA science majors (b; 21 percentage point increase in SYGR) and for first-generation college students with science majors (c; 18 percentage point increase in SYGR).*

These gains are even more substantial for underrepresented groups of students. For AALANA science majors enrolled in high-attrition courses, LA support is associated with an improvement of 21% in SYGR. LA-supported AALANA students had a median SYGR of 81% (71-93%), as compared to the non-LA group of AALANA students, who had a median SYGR of 59% (53-66%). Similarly, for first-generation college students with science majors, the no-LA group had a median SYGR of 65% (60-70%), while the LA group had a median SYGR of 83% (75-91%), for an improvement of 18 percentage points.

## 4 Discussion and Conclusions

We have presented an approach to use Markov chains to estimate SYGR for cohorts of university students. We have used real-world data from RIT to fit this model, and have demonstrated that it faithfully reproduces graduation rate estimates that would be obtained by following a traditional approach. We find that the Markov model yields SYGR estimates with much tighter confidence intervals and lower interannual variability than a traditional approach. We attribute this to the fact that the Markov model leverages all available data, as opposed to only data for cohorts for which at least six years of data are available. This is particularly important for small cohorts of students, which we demonstrated using AALANA students, first-generation college students, and students who experience a specific academic intervention (LA support in high-attrition courses). Confidence intervals for SYGR for underrepresented groups are generally wider, owing to the smaller sample sizes. Even using the Markov model, incorporating all available data, for underrepresented groups who have had LA support in a course, CIs for SYGR can be more than 20 percentage points wide. In these cases, the CIs using a traditional approach are so wide as to lose practical value. Thus, the Markov model for SYGR is particularly valuable in assessing educational program efficacy, especially for underrepresented groups of students.

Graduation rates are frequently used to assess the efficacy of educational programs, but the uncertainties in these graduation rates are not typically addressed or included in these assessments. In this work, we have demonstrated that a Markov model can provide tighter and less variable confidence intervals by incorporating all student data as it becomes available. This is important for academic institutions and programs to provide the most up-to-date information as possible about program efficacy and student outcomes. This work indicates a path forward that addresses the issue that when a university shares graduation rate information, this information is out-of-date as soon as it is posted. Practically none of the students whose data went into the calculation of that graduation rate are still at that university.

Through our demonstration of the utility of the Markov model for small subsets of students, we find that Learning Assistant support in high-attrition gateway science courses is associated with a 9 percentage point improvement in SYGR among LA-supported students, as compared to their non-LA-supported counterparts. We stress that i) the LA and no-LA groups of students are demographically similar, ii) students have no advance knowledge of which sections of a course will be LA-supported when they enroll, and iii) aside from LA support, there are no further systematic differences between LA-supported sections of a course and non-LA-supported sections (e.g., sections use the same syllabus). We acknowledge that our approach cannot directly correct for potential "instructor effects" based on individual instructors or their pedagogical choices. However, our results for the gains in SYGR associated with LA support are consistent from year to year, across all instructors and courses for the high-attrition introductory science courses studied here. Further, the integration of LAs into a course supports a variety of evidence-based pedagogies [6,7], including, for example, more active learning approaches. In light of the demonstrated benefits of active learning [16–18], we interpret our findings as stemming from LA support enabling instructors to implement more evidence-based pedagogies, which in turn leads to improvements in student performance.

These improvements that are associated with LA support are substantially larger for underrepresented groups of students. LA-support is associated with a 21 percentage point improvement in SYGR for AALANA students and an 18 percentage point improvement for first-generation college students. The SYGR estimates for these small subsets of students have wide CIs, making it difficult to disentangle the impacts from an academic intervention such as undergraduate LAs. However, we have demonstrated here that the Markov approach can successfully constrain SYGR estimates for subsets as small as a few dozen students. In addition to the notable improvements to SYGR, LA support is associated with improvements in year-to-year persistence throughout a student's time at RIT (see Supplementary Material). Our analysis indicates that undergraduate LA support in high-attrition introductory science courses may be a potentially fruitful pathway to address persistence and graduation rate gaps between non-underrepresented students and underrepresented students.

## Acknowledgments

We thank the members of the Science and Math Education Research Collaborative at the Rochester Institute of Technology for fruitful conversations. This work was supported in part by a grant from the National Science Foundation, DUE #1757477. This research and use of the longitudinal data set was approved by the Rochester Institute of Technology Human Subjects Research Office (HSRO #01011221; FWA #00000731). The code accompanying this work is available from https://github.com/tonyewong/LA_Markov_analysis. Raw data contains private student information and may not be shared. Processed summary data may be available directly from the authors, subject to Rochester Institute of Technology IRB approval.

13. Goertzen RM, Brewe E, Kramer LH, Wells L, Jones D. Moving toward change: Institutionalizing reform through implementation of the Learning Assistant model and Open Source Tutorials. Phys Rev ST Phys Educ Res. 2011;7: 020105. doi:10.1103/PhysRevSTPER.7.020105

14. Herrera X, Nissen JM, Van Dusen B. Student Outcomes Across Collaborative-Learning Environments. 2018. Available: https://www.per-central.org/items/detail.cfm?ID=14796

15. Van Dusen B, Nissen J. Equity in college physics student learning: A critical quantitative intersectionality investigation. J Res Sci Teach. 2020;57: 33–57. doi:10.1002/tea.21584

16. Deslauriers L, McCarty LS, Miller K, Callaghan K, Kestin G. Measuring actual learning versus feeling of learning in response to being actively engaged in the classroom. Proc Natl Acad Sci. 2019;116: 19251–19257. doi:10.1073/pnas.1821936116

17. Freeman S, Eddy SL, McDonough M, Smith MK, Okoroafor N, Jordt H, et al. Active learning increases student performance in science, engineering, and mathematics. Proc Natl Acad Sci. 2014;111: 8410–8415. doi:10.1073/pnas.1319030111

18. Theobald EJ, Hill MJ, Tran E, Agrawal S, Arroyo EN, Behling S, et al. Active learning narrows achievement gaps for underrepresented students in undergraduate science, technology, engineering, and math. Proc Natl Acad Sci. 2020;117: 6476–6483. doi:10.1073/pnas.1916903117
24

# Supplementary Material
accompanying Improving Models for Student Retention and Graduation using Markov Chains by Tedeschi et al.

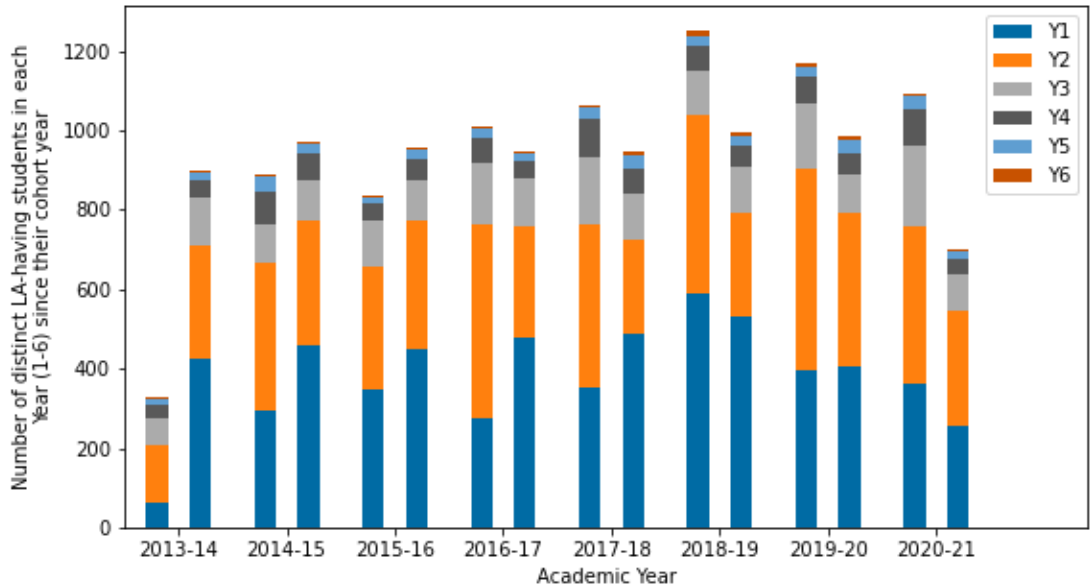

*Figure S1.* Learning assistants consistently affect more than 800 students each semester, with about 20% of these students being in Year 3 or higher. In each pair of stacked bars, the left and right bars represent the fall and spring semesters from that academic year, respectively.

*Table S1.* Numbers of LA-supported and non-LA-supported students and course sections for each high-DFW College of Science course in the semesters from Fall 2013 to Spring 2021. Entries are numbers of students with numbers of course sections in parentheses.

| | | Fa 2013 | Sp 2014 | Fa 2014 | Sp 2015 | Fa 2015 | Sp 2016 | Fa 2016 | Sp 2017 | Fa 2017 | Sp 2018 | Fa 2018 | Sp 2019 | Fa 2019 | Sp 2020 | Fa 2020 | Sp 2021 | Total |
|---|---|---|---|---|---|---|---|---|---|---|---|---|---|---|---|---|---|---|
| College Physics I (PHYS 111) | LA | 0 (0) | 0 (0) | 0 (0) | 0 (0) | 0 (0) | 0 (0) | 79 (1) | 123 (1) | 0 (0) | 0 (0) | 0 (0) | 0 (0) | 0 (0) | 0 (0) | 0 (0) | 0 (0) | 202 (2) |
| | no-LA | 395 (6) | 436 (4) | 308 (6) | 416 (5) | 265 (5) | 411 (5) | 164 (3) | 218 (3) | 267 (4) | 355 (4) | 256 (5) | 387 (4) | 277 (4) | 385 (4) | 249 (3) | 403 (4) | 5192 (69) |
| College Physics II (PHYS 112) | LA | 0 (0) | 122 (1) | 0 (0) | 0 (0) | 0 (0) | 0 (0) | 114 (1) | 102 (1) | 0 (0) | 0 (0) | 0 (0) | 99 (1) | 0 (0) | 0 (0) | 64 (1) | 0 (0) | 501 (5) |
| | no-LA | 137 (2) | 137 (2) | 175 (2) | 167 (2) | 151 (2) | 165 (2) | 30 (1) | 42 (1) | 132 (2) | 146 (2) | 126 (2) | 61 (1) | 121 (2) | 138 (2) | 83 (1) | 146 (1) | 1957 (27) |
| University Physics I (PHYS 211) | LA | 84 (2) | 166 (4) | 42 (1) | 42 (1) | 42 (1) | 123 (3) | 123 (3) | 207 (5) | 210 (5) | 210 (5) | 149 (4) | 123 (3) | 166 (4) | 83 (2) | 129 (4) | 144 (3) | 2043 (50) |
| | no-LA | 213 (6) | 182 (5) | 134 (4) | 304 (8) | 175 (5) | 219 (6) | 93 (4) | 189 (7) | 28 (2) | 214 (7) | 50 (2) | 264 (7) | 75 (2) | 259 (7) | 70 (2) | 250 (8) | 2719 (82) |
| University Physics IA (PHYS 211A) | LA | 0 (0) | 0 (0) | 0 (0) | 0 (0) | 0 (0) | 0 (0) | 42 (1) | 0 (0) | 0 (0) | 0 (0) | 16 (1) | 64 (2) | 17 (1) | 0 (0) | 35 (1) | 82 (2) | 256 (8) |
| | no-LA | 114 (3) | 179 (5) | 111 (3) | 189 (5) | 116 (4) | 258 (7) | 102 (3) | 151 (5) | 112 (3) | 110 (4) | 22 (1) | 42 (1) | 35 (1) | 88 (3) | 0 (0) | 0 (0) | 1629 (48) |
| University Physics II (PHYS 212) | LA | 65 (2) | 81 (2) | 82 (2) | 121 (3) | 69 (2) | 167 (4) | 90 (2) | 80 (2) | 25 (1) | 76 (2) | 115 (3) | 129 (4) | 81 (2) | 190 (5) | 90 (3) | 83 (2) | 1544 (41) |
| | no-LA | 261 (8) | 326 (10) | 333 (9) | 153 (5) | 301 (9) | 171 (6) | 418 (12) | 312 (10) | 420 (13) | 260 (8) | 294 (9) | 144 (4) | 279 (8) | 118 (4) | 261 (7) | 246 (7) | 4297 (129) |
| Precalculus (MATH 111) | LA | 0 (0) | 0 (0) | 36 (1) | 0 (0) | 78 (2) | 0 (0) | 0 (0) | 0 (0) | 0 (0) | 0 (0) | 0 (0) | 0 (0) | 0 (0) | 0 (0) | 0 (0) | 0 (0) | 114 (3) |
| | no-LA | 440 (13) | 216 (6) | 425 (12) | 131 (4) | 426 (12) | 174 (5) | 238 (8) | 102 (4) | 218 (7) | 100 (4) | 211 (6) | 85 (3) | 409 (14) | 97 (3) | 519 (14) | 103 (3) | 3894 (118) |
| Calculus A (MATH 171) | LA | 0 (0) | 0 (0) | 0 (0) | 0 (0) | 0 (0) | 0 (0) | 0 (0) | 0 (0) | 0 (0) | 0 (0) | 268 (7) | 0 (0) | 0 (0) | 0 (0) | 0 (0) | 0 (0) | 268 (7) |
| | no-LA | 348 (9) | 279 (7) | 372 (9) | 301 (8) | 282 (8) | 275 (8) | 244 (7) | 143 (4) | 248 (7) | 132 (4) | 0 (0) | 151 (4) | 333 (9) | 197 (6) | 332 (9) | 248 (6) | 3885 (105) |
| Calculus II (MATH 182A) | LA | 0 (0) | 0 (0) | 0 (0) | 0 (0) | 0 (0) | 0 (0) | 0 (0) | 0 (0) | 0 (0) | 0 (0) | 0 (0) | 0 (0) | 0 (0) | 29 (1) | 0 (0) | 0 (0) | 29 (1) |
| | no-LA | 0 (0) | 165 (5) | 0 (0) | 202 (6) | 35 (1) | 369 (11) | 70 (2) | 269 (8) | 66 (2) | 314 (8) | 77 (2) | 303 (9) | 77 (2) | 228 (6) | 103 (3) | 289 (7) | 2567 (72) |
| Elements of Multivariable Calculus and Differential Equations (MATH 211) | LA | 0 (0) | 0 (0) | 44 (1) | 110 (3) | 144 (4) | 132 (3) | 89 (3) | 119 (3) | 138 (4) | 110 (3) | 146 (4) | 39 (1) | 0 (0) | 102 (3) | 0 (0) | 0 (0) | 1173 (32) |
| | no-LA | 137 (4) | 77 (4) | 85 (3) | 0 (0) | 0 (0) | 0 (0) | 34 (1) | 0 (0) | 0 (0) | 0 (0) | 0 (0) | 70 (2) | 156 (4) | 0 (0) | 154 (4) | 117 (3) | 830 (25) |

*Table S2.* Estimates of confidence intervals for six-year graduation rates (%) are not sensitive to increasing the size of the bootstrap ensembles relative to the main text (1,000 replicates). A traditional calculation for the Fall 2013 cohort is used here as an example.

| Ensemble size | 2.5th percentile | Median | 97.5th percentile |
|---|---|---|---|
| 1000 | 66.4 | 70.6 | 74.8 |
| 2000 | 66.2 | 70.5 | 75.0 |
| 4000 | 66.2 | 70.5 | 75.0 |
| 8000 | 66.2 | 70.6 | 75.2 |

*Table S3.* Learning assistant (LA) support is associated with increases in year-to-year persistence (%) for science majors. Persistence rates are estimated as the Markov transition probability using all available data (Fall 2013 - Summer 2021). Numbers are rounded to the nearest percentage point.

| Transition | No-LA | LA | Difference (LA - no-LA) |
|---|---|---|---|
| Y1→Y2 | 87 | 95 | +9 |
| Y2→Y3 | 91 | 93 | +2 |
| Y3→Y4 | 91 | 93 | +3 |

*Table S4.* Learning assistant (LA) support is associated with increases in year-to-year persistence (%) for AALANA science majors. Persistence rates are estimated as the Markov transition probability using all available data (Fall 2013 - Summer 2021). Numbers are rounded to the nearest percentage point.

| Transition | No-LA | LA | Difference (LA - no-LA) |
|---|---|---|---|
| Y1→Y2 | 84 | 92 | +9 |
| Y2→Y3 | 88 | 92 | +4 |
| Y3→Y4 | 89 | 94 | +5 |

*Table S5.* Learning assistant (LA) support is associated with increases in year-to-year persistence (%) for first-generation science majors. Persistence rates are estimated as the Markov transition probability using all available data (Fall 2013 - Summer 2021). Numbers are rounded to the nearest percentage point.

| Transition | No-LA | LA | Difference (LA - no-LA) |
| --- | --- | --- | --- |
| Y1→Y2 | 86 | 96 | +9 |
| Y2→Y3 | 90 | 93 | +3 |
| Y3→Y4 | 90 | 91 | +1 |